\documentclass[%
 reprint,
 amsmath,amssymb,
 aps,
]{revtex4-2}

\usepackage{graphicx}% Include figure files
\usepackage{dcolumn}% Align table columns on decimal point
\usepackage{bm}% bold math
\usepackage{siunitx}
\begin{document}

\preprint{APS/123-QED}

\title{Counter-propagating spontaneous parametric down-conversion source\\ in lithium niobate on insulator}% Force line breaks with \\
%\thanks{}%

\author{Jost Kellner}
\altaffiliation[These ]{authors contributed equally to this work.}
\author{Alessandra Sabatti}%
\altaffiliation[These ]{authors contributed equally to this work.}
\author{Tristan Kuttner}
\author{Robert J. Chapman}
\author{Rachel Grange}
\email{kellnerj@ethz.ch}

\affiliation{%
 Optical Nanomaterial Group, Institute for Quantum Electronics, Department of Physics, ETH Zurich, CH-8093 Zurich, Switzerland}%

\date{\today}% It is always \today, today,

\begin{abstract}
%Integrated sources of indistinguishable photon pairs are essential for scalable quantum photonic systems. We demonstrate the first integrated counter-propagating photon-pair source on a periodically poled lithium niobate on insulator (LNOI) platform, where the signal and idler photons propagate in opposite directions. The unique phase-matching geometry enables a fixed signal wavelength with a widely tunable idler, as confirmed by sum-frequency generation measurements. We characterize the source indistinguishability through Hong-Ou-Mandel interference in both time and frequency domains, achieving visibilities up to (87.3$\pm$0.6)\%. Spectral purity is assessed via joint spectral intensity measurements and unheralded \( g^{(2)} \) correlations, yielding purities of (96$\pm$4)~\%. Finally, we demonstrate heralded Hong-Ou-Mandel interference between two independent sources, with observed visibilities of $(46\pm6)\%$ for signal and $(71\pm3)\%$ for idler heralding. These results establish counter-propagating photon-pair sources as a highly tunable and pure platform for quantum interference, with direct applicability to multiplexed and reconfigurable quantum photonic networks.
Quantum photonic technologies rely on the ability to generate, manipulate, and interfere indistinguishable single photons on a scalable platform. Among the various approaches, spontaneous parametric down-conversion (SPDC) remains one of the most widely used methods for generating entangled or pure photon pairs. However most integrated SPDC sources relying on co-propagating geometries have a limited purity of heralded photons, or require lossy filtering. Type-2 SPDC processes can produce pure separable photons but typically suffer from lower efficiency and added complexity due to polarisation management.
Here we show the first integrated counter-propagating photon-pair source on lithium niobate on insulator, where signal and idler photons are generated in opposite directions. The counter-propagating geometry leads to spectrally uncorrelated photon pairs without spectral filtering. The joint spectral intensity measurements and unheralded \( g^{(2)} \) correlations, yield purities of (92$\pm$3)\%. Interference between two independent sources achieves heralded visibilities of (71$\pm$3)\%, confirming the scalability of the platform.
These results establish a new route toward integrated, high-purity, and tunable photon sources. The demonstrated counter-propagating geometry offers a scalable solution for quantum photonic networks.
\end{abstract}

\maketitle

%\tableofcontents

\section*{Introduction}

Spontaneous parametric down-conversion (SPDC) is a widely used and well-studied method to generate single-photon sources \cite{review_single_photons}. In this nonlinear optical process, a coherent pump photon is converted into a pair of lower-energy photons, typically termed signal and idler \cite{spdc_prediction}. These photon pairs have become essential tools in a wide range of experiments, from fundamental studies in quantum optics to practical applications such as quantum key distribution \cite{first_demonstration_nonclassic,BB84}. When driven by strong pumps, SPDC can also be used to generate squeezed states, which are critical for quantum computing and quantum metrology \cite{gaussian_boson_sampling,squeezing_in_ligo}.

Early demonstrations used birefringent phase matching in bulk crystals such as barium borate and lithium niobate (LN) \cite{first_spdc,first_shg_ln}.
Later, periodically poled bulk crystals in LN and KTP have been fabricated to tune the features of the source \cite{quasi_phaematching_first}. A major improvement in efficiency was achieved with the development of titanium indiffused waveguides in LN \cite{first_indifused_waveguides,first_qpm_old}, and more recently with the fabrication of high-confinement waveguides in thin films on the lithium niobate on insulator (LNOI) platform \cite{first_waveguide_ln}. These structures enable highly efficient nonlinear processes, including second-harmonic generation, supercontinuum generation, high-speed electro-optic modulators and SPDC \cite{science_review_ln,strong_shg_ring,wang2018integrated,super_continium}.

%In LN, the two most commonly used SPDC processes are Type-0 and Type-II, with co-propagating photons. In Type-0 a transverse electric (TE) field at the fundamental wavelength (typically around \SI{1550}{\nano\meter}) is coupled to a TE field at the second harmonic (\SI{775}{\nano\meter}). In the Type-II process, a TE-polarized photon at \SI{1550}{\nano\meter} and a transverse magnetic (TM)-polarized photon at \SI{1550}{\nano\meter} are coupled to a TM field at \SI{775}{\nano\meter}. Both processes have been demonstrated on the LNOI platform \cite{SPDC_0_LN,xin_spectrally_2022}.

\begin{figure*}
    \centering
    \includegraphics[width=1\linewidth]{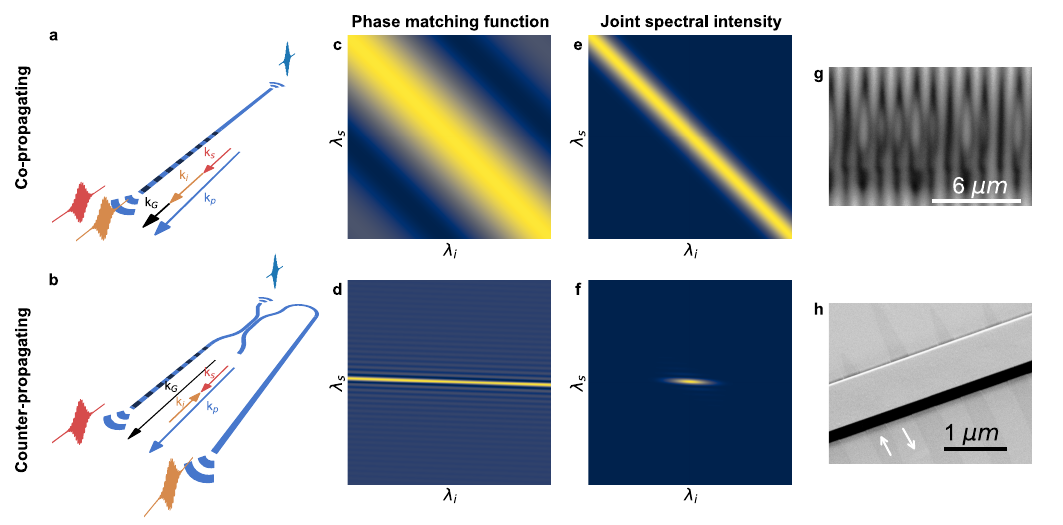}
    
    \caption{\textbf{Spontaneous parametric down-conversion sources in lithium niobate.} 
Schematics of the co-propagating SPDC (\textbf{a}) and counter-propagating SPDC (\textbf{b}) sources, with coloured arrows indicating the directions of the momenta for the pump (blue), signal (red), and idler (orange) photons. The black arrow denotes the momentum given by the periodic poling grating. 
\textbf{c, d}, Phase-matching functions for the co- and counter-propagating configurations, respectively. 
\textbf{e, f}, Joint spectral intensities resulting from the overlap of the pump spectrum with the corresponding phase-matching function for the co- and counter-propagating cases. 
\textbf{g}, Two-photon microscope image of the poled thin-film before waveguide fabrication.
\textbf{h}, Scanning electron microscope image of the fabricated waveguide with visible grey shades of the inverted domains. White arrows indicate the orientation of the domains.}

    \label{fig:fig1}
\end{figure*}

Generating spectrally and spatially separable photons with high purity is a key goal for single photon pair sources \cite{review_single_photons,PhysRevA.98.053811}. High purity enables heralding of one photon without perturbing the quantum state of the other, which is essential for multi-photon interference and the generation of complex, high-dimensional quantum states \cite{HeraldedPhotons}. The typical co-propagating Type-0 SPDC produces photon pairs in a single spatial mode (see Fig.~\ref{fig:fig1}a). Co-propagating Type-0 SPDC has a phase matching function shown in Fig.~\ref{fig:fig1}c, which produces maximally entangled photon pairs in the frequency domain, as shown in the JSI in Fig.~\ref{fig:fig1}e \cite{SPDC_0_LN,luo_counter-propagating_2020}. In contrast, Type-II processes can achieve nearly uncorrelated photon pairs with high spectral purity \cite{xin_spectrally_2022,kuttner2025scalablequantuminterferenceintegrated}. However, the Type-II configuration harvests a smaller component of the nonlinear tensor in the LN, thus suffering from reduced efficiency. Another disadvantage is the need to manage orthogonal polarisations. Type-0 benefits from the strongest nonlinear tensor component and the use of a single polarisation, but requires filtering to achieve high purity, which reduces both generation and detection efficiencies \cite{filtering_effincey_purity}.

A promising approach to combine the benefits of both processes was proposed by Booth et al.~\cite{OG_counter_prop}. By phase-matching two photons propagating in opposite directions, one can leverage the strong nonlinear tensor of LN while generating spectrally separable photons in distinct spatial modes. Fig.~\ref{fig:fig1}b illustrates the two generated photons (orange and red) exiting the waveguide through different ports. Due to the favourable dispersion properties of this configuration, the generated photon pair is intrinsically spectrally uncorrelated and exhibits high purity \cite{Pure_counterprop}. The phase-matching function and JSI for such a process are shown in Fig.~\ref{fig:fig1}d and Fig.~\ref{fig:fig1}f, respectively. This method was first demonstrated in bulk periodically poled KTP \cite{canalias_mirrorless_2007} and later on in indiffused waveguides with fifth and third order poling \cite{luo_counter-propagating_2020, PPKTP_5th, PPKTP_5th, Pure_counterprop} and more recently on z-cut LNOI for the generation of symmetric second-harmonic \cite{yang_symmetric_2024}. To the best of our knowledge, a counter-propagating SPDC source has never been reported in an integrated platform, and there has not been any implementation of this phase matching process on x-cut LNOI. Demonstrating this process on the x-cut platform, which is widely used for integrated photonics, would enable seamless integration with modulators and other established components.

Here, we present the first fully integrated counter-propagating SPDC source on the lithium niobate on insulator x-cut platform. We characterise the spectral properties of the source and compare the measured phase-matching function with numerical simulations. We demonstrate broad tunability and verify the quantum nature of the source using Hong–Ou–Mandel interference yielding a visibility of (87.3$\pm$0.6)\% for continuous wave operation. We characterize the source joint spectral intensity and with a unheralded $g^{(2)}$ measurement and confirm a high purity of $(92\pm3)\%$, obtained while pumping the source with a pulsed laser with a bandwidth of \SI{1.1}{nm}. Finally, we interfere photons from two independent sources and confirm the high spectral purity of the heralded single photons. By heralding the two idler photons of the two sources, we achieve a Hong-Ou-Mandel interference with a visibility of $(71\pm3)\%$.

\section*{Results}

\begin{figure*}
    \centering
    \includegraphics[width=1\linewidth]{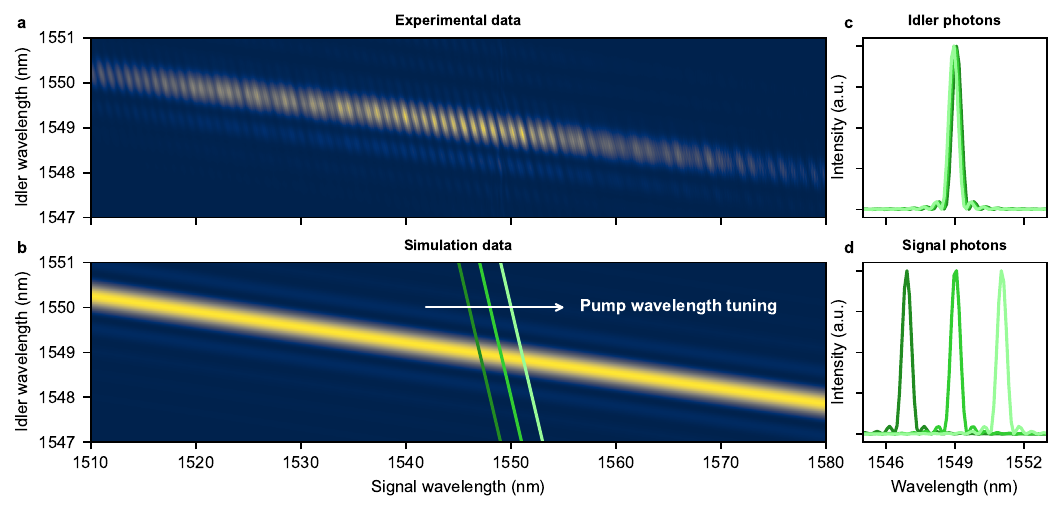}
    \caption{\textbf{Phase-matching function characterisation.} 
\textbf{a}, Experimentally measured sum-frequency generation spectrum. 
\textbf{b}, Simulated SFG spectrum, overlaid with energy conservation lines corresponding to three different continuous-wave pump wavelengths. 
\textbf{c}, Retrieved idler photon spectra for the three pump wavelengths, showing minimal variation. 
\textbf{d}, Corresponding signal photon spectra, exhibiting strong dependence on the pump wavelength.}

    \label{fig:fig2}
\end{figure*}

\subsection{Source Design and Classical Characterisation}
The source design incorporates three grating couplers for the in- and out-coupling of light. Two of these couplers are optimised for operation near \SI{1550}{nm}, corresponding to the signal and idler photons, while a third is designed for the pump field at \SI{775}{nm}. Light entering through the pump grating coupler is routed into a directional coupler, which functions as a wavelength demultiplexer by allowing the pump to pass through the same arm while redirecting light in the C-band. Photon pairs are subsequently generated in the periodically poled region: the signal photon propagates forward and exits the chip directly, whereas the idler photon travels backwards and is routed out via the wavelength demultiplexer and its corresponding grating coupler. A schematic of the source is shown in Fig.~\ref{fig:fig1}b. Since the momenta of the signal and idler photons cancel each other, the poling momentum must compensate for the full pump momentum. Due to the technical challenges associated with sub-micrometre poling periods, we employ a third-order poling scheme with a poling period of \SI{1.18}{\micro m}. The corresponding period for first-order poling would be approximately \SI{400}{nm}. Phase matching can still be achieved using higher-order QPM, provided that the order is odd \cite{high_order_PPLN}. For poling with order \textit{n}, the efficiency scales with $\frac{1}{n^2}$ in respect to the first order efficiency. For a more detailed discussion, see the Supplementary information. A two-photon microscope image of the poled region is presented in Fig.~\ref{fig:fig1}g before fabrication. Fig.~\ref{fig:fig1}h shows a scanning electron microscope image of the fabricated waveguides. The inverted domains are visible thanks to the contrast in the electronic signal.

% here we discuss the phase matching function, sfg results and the nice features of the tunabiliy
We first characterised the source by measuring the phase-matching function (PMF) via sum-frequency generation. Two tunable lasers, operating near \SI{1550}{nm}, were coupled into the device through the grating couplers corresponding to the signal and idler modes. Their nonlinear interaction generates light at the pump wavelength, which is detected using a silicon photodetector. By raster-scanning the input wavelengths, we map out the PMF. The experimentally measured function is shown in Fig.~\ref{fig:fig2}a, and is in good agreement with the simulated PMF in Fig.~\ref{fig:fig2}b. Minor oscillations observed in the pattern are attributed to weak Fabry–Pérot resonances caused by reflections of \SI{775}{nm} light within the device.

\subsection{Continuous-wave pump regime}

Under continuous-wave (CW) pumping, the counter-propagating source reveals a unique spectral property. In Fig.~\ref{fig:fig2}b, three green lines represent energy conservation for different pump wavelengths. By projecting the product of the phase-matching function and the pump field (i.e., the energy conservation line) onto the signal or idler axis, we retrieve the individual spectral intensity profiles of the generated photons. Due to the flat slope of the PMF, the spectrum of the idler photon remains nearly invariant to pump tuning (Fig.~\ref{fig:fig2}c), whereas the signal spectrum exhibits strong sensitivity to the pump wavelength (Fig.~\ref{fig:fig2}d). This asymmetry suggests promising applications: the idler photon can remain fixed within the telecommunications C-band, while the signal can be spectrally tuned to interface with different systems. With careful adjustment of the pump wavelength, the degree of spectral overlap between the signal and idler photons can be controlled, enabling their emission to be either indistinguishable or distinguishable.

\begin{figure*}
    \centering
    \includegraphics[width=1\linewidth]{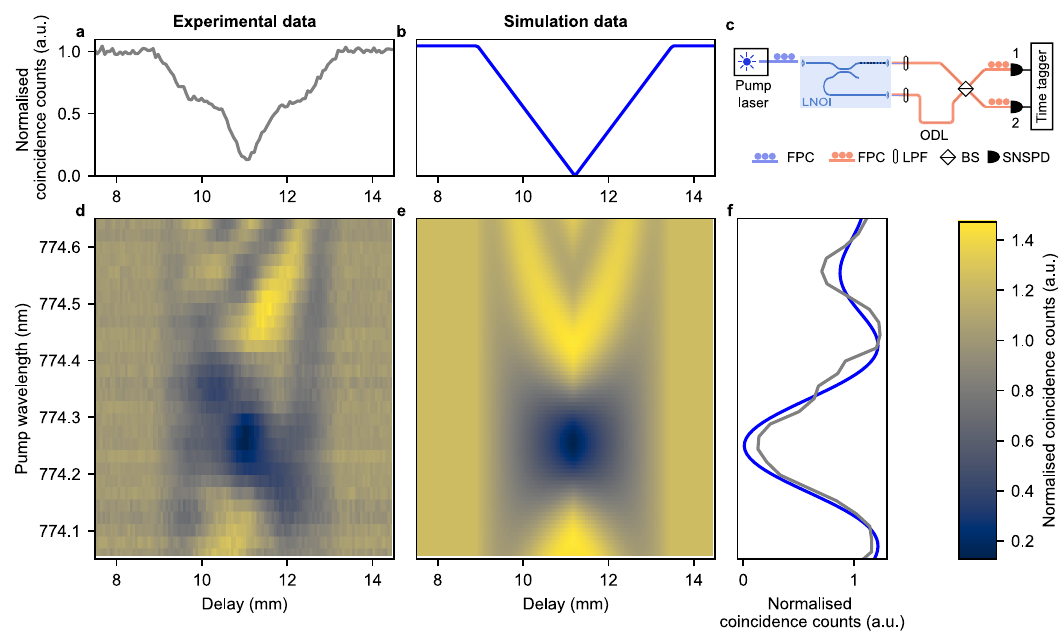}
    \caption{\textbf{Hong–Ou–Mandel interference in time and frequency domains.} 
\textbf{a}, HOM interference pattern measured at the pump wavelength where the signal and idler photons are indistinguishable. 
\textbf{b}, Expected HOM dip for an ideal system. 
\textbf{c}, Schematic of the measurement setup. A tunable continuous-wave laser is coupled into the chip via a fibre polarisation controller (FPC). The generated signal and idler photons are out-coupled from the chip; the idler photon passes through a tunable optical delay line (ODL). Both photons interfere on a beam-splitter (BS) and are detected by superconducting nanowire single-photon detectors (SNSPDs), with coincidence events recorded using a time tagger. 
For each pump wavelength, a HOM interference pattern is measured by varying the photon delay, resulting in the experimental map shown in \textbf{d} and the corresponding theoretical map in \textbf{e}. \textbf{f}, Spectral-domain projection of the HOM interference, obtained by slicing the maps in \textbf{d} and \textbf{e} at zero temporal delay (\SI{11}{mm}. The simulation is plotted in blue and the measured in gray.
}
    \label{fig:fig3}
\end{figure*}

Hong–Ou–Mandel (HOM) experiments involve a single source producing indistinguishable photons. By varying the temporal offset between the photons, one can observe interference effects such as photon bunching \cite{hong_measurement_1987}. In our case, we use the counter-propagating on-chip source for generating both the signal and idler photons, which are interfered with an off-chip beam-splitter, as schematically illustrated in Fig.~\ref{fig:fig3}c. To change the temporal offset we use a fibre-coupled optical delay line which changes the path of the idler photon by adjusting the length. The photon pair rate in our experiment is 20~kHz, with a measured coincidence-to-accidental ratio of 3000 at a pump power of \SI{7}{mW}.

To observe HOM interference, the pump wavelength is first tuned to the value at which the signal and idler photons become spectrally indistinguishable. This optimal condition can be identified from the SFG map shown in Fig.~\ref{fig:fig2}a. Given the sinc-shaped spectral profiles of the photons, we expect a triangular-shaped HOM dip as shown in the simulation (Fig.~\ref{fig:fig3}b) \cite{branczyk_hong-ou-mandel_2017}. The experimentally observed dip is presented in Fig.~\ref{fig:fig3}a, with a visibility of (87.3$\pm$0.6)\%. Deviations from the ideal dip shape are attributed to spectral imperfections in the source, likely arising from slight variations in the poling period. The limited visibility can also be explained by reflective effects associated with the grating couplers (see supplementary information). %While these couplers are optimised for out-of-plane transmission, they approach a stop-band regime that reflects a portion of the light back into the waveguide. This effect is supported by measurements of bidirectional photon arrival histograms (\textcolor{red}{see Suplementary}). 
%In contrast to controlling indistinguishability via temporal delay, we also explore spectral distinguishability  
The indistinguishability can be controlled not only using time as a turning knob, but also in the frequency domain, by varying the pump wavelength while keeping the temporal delay fixed. The principle behind spectral tuning exploits the effect represented in Fig.~\ref{fig:fig2}b–d. The bunching arises from the change of the signal spectrum, while the idler remains fixed. We present this measurement in a two-dimensional map of HOM interference, with the pump wavelength on the vertical axis and the idler path delay on the horizontal axis. For each pump wavelength, a normalised HOM dip is measured. By extracting a vertical slice at zero delay (in our case \SI{11}{mm}), we obtain a HOM dip in frequency space.
\begin{figure*}
    \centering
    \includegraphics[width=1\linewidth]{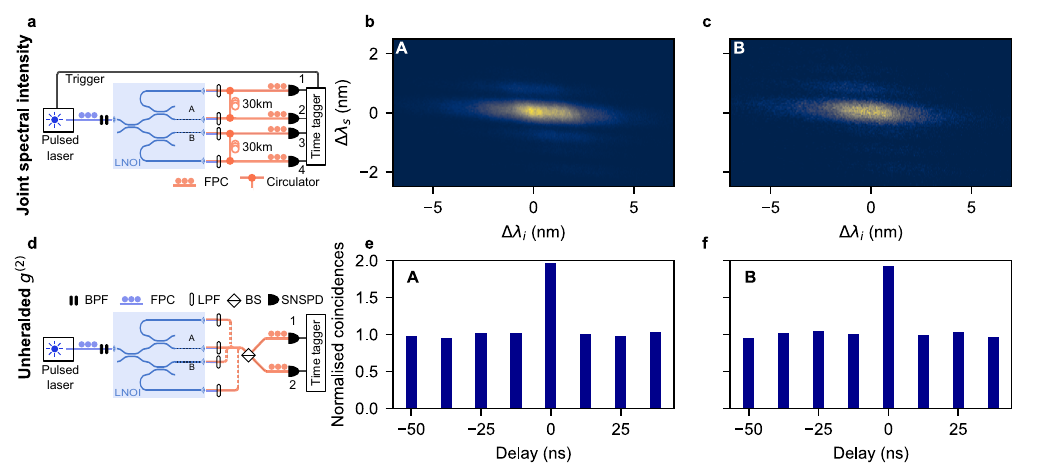}
    \caption{\textbf{Joint spectral intensity measurement and unheralded $g^{(2)}$ analysis.} 
\textbf{a}, Experimental setup for measuring the joint spectral intensity (JSI). A pulsed laser at \SI{775}{nm} with a spectral bandwidth of \SI{1.1}{nm} is coupled to the chip. The signal and idler photons are sent through a \SI{30}{km} dispersion module, and their arrival times are recorded to reconstruct the JSI. 
\textbf{b}, \textbf{c}, Measured JSIs for sources A and B, respectively. 
\textbf{d}, Setup for the unheralded $g^{(2)}$ measurement, where only the signal (or idler) photon is analysed via a balanced beam-splitter and two detectors. 
\textbf{e}, \textbf{f}, Extracted purities for sources A and B, respectively, based on unheralded $g^{(2)}$ measurements, with reported values of (96$\pm$4)~\% and (92$\pm$3)~\%. 
Abbreviations: FPC, fibre polarisation controller; BPF, bandpass filter; LPF, low-pass filter; BS, beam-splitter; SNSPD, superconducting nanowire single-photon detector.}

    \label{fig:fig4}
 \end{figure*}
The measured interference map is shown in Fig.~\ref{fig:fig3}~d and compared with a simulated map in Fig.~\ref{fig:fig3}e. Asymmetries in the interference pattern suggest residual imperfections in the poling, which affect the spectral symmetry of the photons. The corresponding spectral slices from the experimental and simulated data are plotted in Fig.~\ref{fig:fig3}~f. In this case, the HOM dip follows a sinc-like profile, in contrast to the triangular dip observed in the time-delay domain. We report the same (87.3$\pm$0.6)\% visibility for both measurements, as they correspond to projections through the same central region of the map. We note that both the measured and simulated normalised coincidence maps exhibit values exceeding 1 at certain positions. Values above 1 indicate anti-bunching behaviour, where the probability of the photons exiting different ports of the beam-splitter is higher than that expected from random splitting. In this normalisation, a value of 1 corresponds to classical random splitting, while 0 corresponds to perfect photon bunching at the beam-splitter. Values above 1 are also reported in top-hat like photon spectra, typically resulting from band-pass filtering, where the HOM dip follows a sinc-shape \cite{Maeder2024}.

\subsection{Pulsed pump regime}
 
As discussed above, the source shows promising performance under continuous-wave operation. However, one of its key advantages is the generation of heralded single photons. In the following, we investigate the purity and spectral characteristics of the source using three complementary methods. For these measurements, we use a mode-locked titanium-sapphire laser operating at a repetition rate of \SI{80}{MHz} centred at \SI{774}{nm}, spectrally filtered to a bandwidth of \SI{1.1}{nm}. The full details of the setup are described in the method section. We characterise two independent on-chip sources by measuring their joint spectral intensity (JSI), their unheralded second-order correlation function $g^{(2)}$, and finally, their indistinguishability in a two-source heralded Hong–Ou–Mandel experiment.

The first method to quantify the spectral purity is based on the measured joint spectral intensity. Since this measurement does not access phase information, the calculated purity reflects only the amplitude correlations. The JSI is measured using the trigger signal from the laser and two single-photon detectors connected to a time tagger. To obtain spectral resolution, the signal and idler photons are routed through circulators and a dispersion module equivalent to \SI{30}{km} of fibre (Fig.~\ref{fig:fig4}~a). The trigger provides a temporal reference, and the relative arrival times of the photons are converted to wavelengths using the known dispersion profile. The resulting JSIs of the two sources are shown in Fig.~\ref{fig:fig4}~b and c. Note that the absolute central wavelengths cannot be retrieved from this measurement but are inferred from the previously measured SFG spectra. We calculate purities of 83.1\% and 84.5\% for sources A and B, respectively. The purity values are computed by applying singular value decomposition to the extracted JSAs from the measured JSIs. %Singular value decomposition is applied to the measured JSIs extracted JSAs to compute the purity values.  

The second method involves direct measurement of the unheralded second-order correlation function $g^{(2)}$. Using the same pump configuration, we isolate one of the sources and collect only the idler photon (Fig.~\ref{fig:fig4}~d). A balanced directional coupler splits the photon stream to two detectors, and the coincidence histogram is recorded. The central peak reflects the photon statistics, while the side peaks are normalised to unity. A higher central peak corresponds to higher purity. We perform this measurement on the idler photons from sources A and B. The results are shown in Fig.~\ref{fig:fig4}~e–f, yielding purities of (96$\pm$4)~\% and (92$\pm$3)~\%, respectively. The discrepancy between the values obtained from JSI and $g^{(2)}$ measurements is attributed to uncertainties in the limited spectral range of the JSI technique and potential noise in the correlation measurements. This experiment is performed in a non-degenerate configuration because small reflections at the output grating couplers can cause signal photons to exit through the idler port, as observed in the CW pump regime. To suppress any spurious signal photons, we introduce a \SI{12}{nm} bandpass filter centered at \SI{1550}{nm}.

\begin{figure}
    \centering
    \includegraphics[width=1\linewidth]{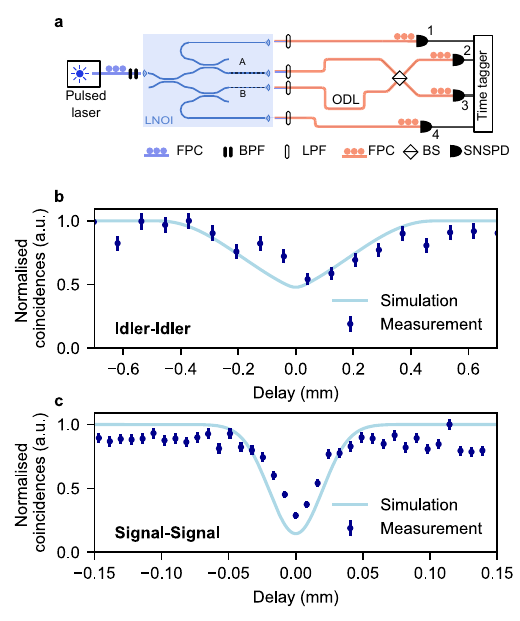}
    \caption{\textbf{Heralded two-source Hong-Ou-Mandel interference.} 
\textbf{a}, Experimental setup for heralded Hong-Ou-Mandel (HOM) interference between either the signal–signal or idler–idler photon pairs from two independent sources on the chip. A fibre polarisation controller (FPC) prepares the pump, which is coupled to the chip. Signal and idler photons are separated and filtered with bandpass filters (BPF) and low-pass filters (LPF), and directed through a balanced beam-splitter (BS) for interference. Coincidences are measured using superconducting nanowire single-photon detectors (SNSPD). 
\textbf{b}, HOM interference dip for the heralded idler–idler case: measured data (dark blue) and corresponding simulation (light blue). 
\textbf{c}, HOM interference dip for the heralded signal–signal case: measured data (dark blue) and simulation (light blue).}

    \label{fig:fig5}
 \end{figure}

In a final characterisation step, we simultaneously operate the two integrated sources on the chip to demonstrate their mutual interference. The pump preparation is performed as described previously, using an average power of \SI{2}{mW}. Both signal and idler photons are collected off-chip, and interference is studied by heralding either the signal or the idler photons. The corresponding heralded idler or signal photons are sent through an optical delay line and interfered with a balanced fibre directional coupler. The interfering photons are detected using two single-photon detectors, while the heralding photons are monitored on a separate pair of detectors. By scanning the optical delay line and recording four-fold coincidences, we extract the Hong–Ou–Mandel (HOM) interference dip for both scenarios. The experimental setup is illustrated in Fig.~\ref{fig:fig5}~a.

The measured and simulated HOM dips for the idler and signal photon pairs are shown in Fig.~\ref{fig:fig5}~b and Fig.~\ref{fig:fig5}~c, respectively. We observe a visibility of $(46\pm6)\%$ for the idler pair and $(71\pm3)\%$, for the signal pair, while simulations predict a maximum visibility of 88\%. Notably, the difference in visibility between the signal and idler photon pairs is already present in the simulation and confirmed experimentally. This reduced visibility for the idler photons arises from a mismatch in the phase-matching conditions between the two sources. Specifically, source A has a slightly shifted phase matching function in comparison to source B, which affects the idler photon spectrum in particular. As discussed in Fig.~\ref{fig:fig2}, the signal spectrum closely follows the pump wavelength and bandwidth, whereas the idler spectrum is more strongly determined by the poling period. Introducing local heaters near one of the sources could compensate for these differences by fine-tuning the effective refractive index and, hence, the phase-matching condition, thereby equalising the spectral properties and improving the interference visibility.

\section*{Discussion and conclusion}

In this work, we build on the promising properties of counter-propagating photon-pair sources \cite{luo_counter-propagating_2020,canalias_mirrorless_2007} and, to the best of our knowledge, demonstrate the first integrated implementation of such a source. To characterise its unique features, we performed sum-frequency generation measurements, which reveal the distinct behaviour of the phase-matching function: while the idler photon exhibits tunability with the pump wavelength, the signal photon remains spectrally stable. This asymmetry in spectral behaviour opens up novel opportunities for spectral engineering and hybrid quantum systems.

Furthermore, we verified the high indistinguishability of signal and idler photons by performing Hong–Ou–Mandel (HOM) interference measurements in both the temporal and spectral domains. The spectral tunability of one of the two photons enabled the demonstration of HOM interference beyond the conventional time-delay-based approach with a visibility of (87.3$\pm$0.6)\%, highlighting an additional degree of control.

To further investigate the spectral properties, we measured the joint spectral intensity (JSI) for two independent sources on the same chip, performing for the first time this characterisation for an integrated counter-propagating platform. The extracted purities from the JSI were 83.1\% and 84.5\%. These values were independently confirmed via unheralded $g^{(2)}$ measurements, yielding comparable results and supporting the high spectral purity of the generated photons.

Finally, we leveraged these two sources to demonstrate heralded Hong–Ou–Mandel interference between signal and idler photons. The observed visibilities of $(71\pm3)\%$ underscore the suitability of the source for quantum interference experiments and pave the way toward integrated, multi-photon quantum photonics based on counter-propagating geometries.

In conclusion, our work establishes counter-propagating integrated sources as a powerful and tunable platform for high-purity quantum photonics, with immediate applications in quantum interference and entanglement experiments, and promising prospects for scalable quantum photonic networks. Looking ahead, implementing this type of source with first-order poling around \SI{400}{nm} could lead to significantly higher efficiency and reduced pump power requirements. Moreover, the successful demonstration of integrated counter-propagation provides a basis for more advanced on-chip devices, such as mirror-less optical parametric oscillators pumped by continuous-wave light, or two-mode squeezers for quantum sensing applications.

\section*{Methods}

% \textbf{Device design.}The counterpropagating SPDC source is defined by a periodically polled waveguide sandwiched between one grating coupler (GC) for \SI{1550}{nm} in and out-coupling and a directional coupler as presented in Fig.1. The directional coupler (DC) length is designed such that it functions as a wavelength de-multiplexer. The light around \SI{775}{nm} is coupled to the poled waveguide via the bar port of the DC and the bar port is connected to a second GC for \SI{1550}{nm}. With this design, the source can be either probed by two \SI{1550}{nm} lasers for classical linear and nonlinear characterisation or by a \SI{775}{nm} laser for quantum experiments. The related measurement setups are presented in the related experiment section of this paper. 

\textbf{Device fabrication.} The circuit is realised using the LNOI platform, with a \SI{5}{\%} MgO doped LN film whose thickness is \SI{300}{nm}, and with a \SI{2}{\um} thick buried oxide. The periodic poling is realized prior to waveguide etching, using \SI{100}{nm} thick Ti electrodes for the high voltage pulse application. The electrodes are patterned by electron-beam lithography (EBL), electron-beam evaporation and liftoff. A photoresist that serves as an electrical insulation layer is applied to the chip during the poling process. After the removal of the photoresist and of the poling electrodes, the waveguides are patterned into a FOX16 mask with another EBL step. The film is etched by \SI{200}{nm} using an Argon ion etching process via ICP-RIE \cite{FabPaper}. The LN redeposited during the etching is removed with an RCA-SC1 cleaning step, followed by a BHF dip to remove the remaining mask.

\textbf{Experimental setup.}  
The pulsed pump laser is a Coherent Chameleon Ultra with a repetition rate of \SI{80}{MHz} and a bandwidth of \SI{8}{nm}. It is spectrally filtered down to \SI{1.1}{nm} using two cascaded bandpass filters (Thorlabs FBH780-10). The laser is coupled into a polarisation-maintaining (PM) fibre using a collimator and a half-wave plate to set the polarisation. Input coupling to the chip is achieved by aligning a cleaved single-mode fibre above the grating coupler at the appropriate angle. For output coupling, a fibre array (Oz Optics) is used. PM long-pass filters (Oz Optics) are placed after the chip to remove residual pump light.To ensure polarisation matching between the photons and the detectors, fibre polarisation controllers (FPCs) are placed in front of each superconducting nanowire single-photon detector (SNSPD). The eight-channel SNSPD system is provided by Single Quantum. Detection events are recorded using a Swabian Ultra time tagger. For Hong-Ou-Mandel experiments, a fibre beam splitter (Thorlabs PN1550R5A2) is used for the interference. For joint spectral intensity measurements, a dispersion module (Tellabs 82.71532-30-R5) equivalent to \SI{30}{km} of fibre is employed.

\section*{Author contributions}
J.K. and A.S. contributed equally to this work. J.K., A.S., and R.J.C. conceived the idea. A.S. and J.K. designed the chip. A.S. fabricated the sample and performed the device-level characterisation. J.K. performed the optical characterisation and optical measurements with inputs form T.K. The manuscript was written by J.K. and A.S. with contributions from all authors. R.G. supervised the project.

\section*{Competing interest}
The authors declare no competing financial or nonfinancial interests.

\begin{acknowledgments}
We acknowledge support for the fabrication and the characterisation of our samples from the Scientific Center of Optical and Electron Microscopy ScopeM and from the cleanroom facilities BRNC and FIRST of ETH Zurich and IBM Ruschlikon. R.J.C. acknowledges support from the Swiss National Science Foundation under the Ambizione Fellowship Program (Project Number 208707) R.G. acknowledges support from the European Space Agency (Project Numbers 4000137426 and 4000136423), the Swiss National Science Foundation under the Bridge Program
(Project Number 194693) and the Sinergia Program (Project Number CRSII5 206008). We thank Chiprin Costea for contributing to some of the preliminary experiments.
\end{acknowledgments}

\section*{Data availability statement}
Data supporting the findings of this study are available within the article and the Supplementary Material. Raw data and analysis code are available from the corresponding author upon reasonable request.\\

\newpage

%\nocite{*}

\bibliography{main}% Produces the bibliography via BibTeX.

\end{document}